\documentclass[%
 amsmath,
 amssymb,
prb,
twocolumn]{revtex4-2}

\usepackage{graphicx}
\usepackage{dcolumn}
\usepackage{bm}
\usepackage{color}
\usepackage{multirow}
\usepackage{xtab,afterpage,longtable}

\DeclareUnicodeCharacter{2212}{-}

\begin{document}


\title{A Python code for calculating the mean-value (Baldereschi's) point for any crystal structure}

\author{Vladan Stevanovi\'c}
\email{vstevano@mines.edu}
\affiliation{Colorado School of Mines, Golden, CO 80401,USA}

\date{\today}

\begin{abstract}
A python code (\texttt{mvp.py}) is presented for computing the mean-value point (MVP) in the Brillouin zone first introduced by Baldereschi \cite{Baldereschi_PRB:1973}. The code allows calculations of the MVP for any input crystal structure. Having MVP allows approximating the Brillouin zone integrals of relatively smooth, periodic functions defined in the reciprocal space by the value of the same function at only one, mean-value, $\mathbf{k}$-point. This approximation decreases computational cost at a relatively small decrease in accuracy. The MVP coordinates for the 14 Bravais lattices are evaluated and the underlying theory is discussed. 
\end{abstract}

\maketitle

\section{Introduction}
%
In his original paper Baldereschi defined the mean-value point in the Brillouin zone as: ``the point such that the value which any given periodic function of wave vector assumes at this point is an excellent approximation to the average value of the same function throughout the Brillouin zone'' \cite{Baldereschi_PRB:1973}. The conditions that define the mean-value point (MVP) follow from the expansion of any periodic function defined on the $\mathbf{k}$-space into plane waves:
\begin{equation}\label{eq:expansion}
f(\mathbf{k}) \, = \, \sum_{\mathbf{R}} \, F(\mathbf{R}) \, e^{i \, \mathbf{R} \, \mathbf{k}},
\end{equation}
where $f$ stands for a periodic function of the wave vector $\mathbf{k}$, the $\mathbf{R}$ represent vectors of the direct (Bravais) lattice, which take the role of the wave vectors in the above expansion, and $F(\mathbf{R})$ are the Fourier components of the function $f$. Generally, the function $f$ needs to fulfill Dirichlet conditions for the expansion from eq.~\eqref{eq:expansion} to converge to $f$. 

From eq.~\eqref{eq:expansion} it is easy to show that the integral of $f$ over the entire Brilloiun zone equals the $F(\mathbf{R}=0)$ term in the expansion multiplied by the volume of the Brilloiun zone ($(2\pi)^3/V$). Hence, if there were a point $\mathbf{k}=\mathbf{k}_\mathrm{o}$ at which the sum of all terms in the expansion above, excluding the $\mathbf{R}=0$ term, vanishes, the value $f(\mathbf{k}_\mathrm{o}) \, (2\pi)^3/V$ would be exactly equal to the integral of $f(\mathbf{k})$ over the entire Brillouin zone. The existence of $\mathbf{k}_\mathrm{o}$ is guaranteed by the mean-value theorem of the integral calculus, but its coordinates are dependent on the function $f$. The mean-value point from Baldereschi's work is a point defined by the symmetry of the crystal that is independent of $f$, at which the value of a function $f$ is an ``excellent'' approximation of its integral over the Brillouin zone.

Hence, approximating the integral of $f(\mathbf{k})$ over the entire Brillouin zone can be reduced to finding $f(\mathbf{k}_\mathrm{o})$, provided that $\mathbf{k}_\mathrm{o}$ is known. In the original work of Baldereschi \cite{Baldereschi_PRB:1973}, the coordinates of $\mathbf{k}_\mathrm{o}$ are evaluated for the three cubic Bravais lattices, face centered cubic (fcc), body centered cubic (bcc), and simple cubic (sc). Subsequently, the mean-value point has been found for some non cubic lattices (see for example \cite{Bashenov_PSSb:1977}). However, to the best of my knowledge, the mean-value point has not been evaluated for all 14 Bravais lattices, and a general, stndalone code for doing so for any input crystal structure is not available. Herein, both the \texttt{mvp.py} code and the resulting MVPs for the 14 Bravais lattice are presented, and various aspects of the procedure are discussed.

\section{Problem setup and code description}
%
The key to evaluating the MVP for a given crystal structure is the requirement that the sum of all terms in eq.~\eqref{eq:expansion} for $\mathbf{R} \neq 0$ vanishes at some $\mathbf{k} = \mathbf{k}_\mathrm{o}$. These conditions provide a system of equations whose solution(s) define $\mathbf{k}_\mathrm{o}$. For the purpose of finding an approximate, function independent $\mathbf{k}_\mathrm{o}$ the sum from eq.~\eqref{eq:expansion} can be rewritten in a more suitable form using the symmetry of the crystal. As already discussed by Baldereschi in Ref.~\cite{Baldereschi_PRB:1973} one can assume, without any loss of generality, that all $F(\mathbf{R})$ are equal for $\mathbf{R}$ vectors that are connected by the point group symmetry operations, that is, that belong to the same star of $\mathbf{R}$. One can then write:
\begin{equation}\label{eq:expansion_2}
f(\mathbf{k}) \, = \, \sum_{s} F_s \sum_{n_s} \, e^{i \, \mathbf{R}_{sn_s} \, \mathbf{k}} \, = \, \sum_{s} F_s \, W_{s}(\mathbf{k}),
\end{equation}
where $s$ goes over all stars of $\mathbf{R}$, $n_s$ counts all the lattice vectors within a given star, and $W_{s}(\mathbf{k})$ are the ``symmetrized'' plane waves belonging to the $\Gamma_1$ (fully symmetric) irreducible representation of the point group.  

Hence, the $f$-independent conditions that determine $\mathbf{k}_\mathrm{o}$ are that all $W_{s}(\mathbf{k}_\mathrm{o})=0$ for any $s>0$. This is, of course, too much to ask and this system of equations does not have a solution.  However, two simplifications can be made. First, one can assume that $F_s$ decrease sufficiently fast with $s$, so that after a relatively small number of terms the non-zero value of $W_s$ becomes irrelevant. Second, even if the number of zero terms is smaller than three, which is necessary for all three components of $\mathbf{k}_\mathrm{o}$ to be uniquely determined, the values can be chosen so to minimize the first nonzero $W_s(\mathbf{k})$. These two assumptions allowed Baldereschi to evaluate the coordinates of the mean-value points for the three cubic systems (sc, bcc, fcc).

The \texttt{mvp.py} code follows the outlined protocol. For any given crystal structure, a set of lattice vectors $\mathbf{R}=n_1\mathbf{a}_1 + n_2\mathbf{a}_2 + n_3\mathbf{a}_3$ is constructed. Here $\mathbf{a}_1,\mathbf{a}_2,\mathbf{a}_3$ are the unit cell vectors and the integers $n_1,n_2,n_3$ are chosen to cover a range of values so that a sufficiently large number of $\mathbf{R}$s is produced; large enough to enclose the first 4 stars of $\mathbf{R}$ ($s=0,1,2,3$). The number of $\mathbf{R}$ vectors is controlled by the \texttt{R\_range} input integer variable ($-$ \texttt{R\_range} $\leq n_1,n_2,n_3 \leq $ \texttt{R\_range}), whose default value is 4. All results presented in this paper are obtained with \texttt{R\_range} = 4. 

Then the point group symmetry operations are found with the help of \texttt{spglib} library \cite{spglib}. For this purpose the tolerance factors for the \texttt{spglib} are set to \texttt{symprec} = 1e-02 {\AA} (length) and \texttt{angle\_tolerance} = $-1.0$ (angles). The full set of orthogonal transformations (rotations and mirror planes) including those that are parts of screw axes and glide planes is used for the classification of $\mathbf{R}$ vectors into stars. Conveniently, the \texttt{spglib} library provides a separation of the of the space group operations into orthogonal parts (labeled ``rotations'') and associated translational parts (fractional translations for the glide planes and screw axes). The \texttt{mvp.py} code takes the entire set of orthogonal parts, and uses that set to classify $\mathbf{R}$ into stars. 
 
In the next step, the $\mathbf{R}$ vectors are classified into classes of symmetry equivalent members forming the stars. This is done by applying the entire set of orthogonal transformations successively on $\mathbf{R}$ vectors and grouping them with those they transform into. The equality of a given $\hat{O}\mathbf{R}$ and some $\mathbf{R}'$ is determined using the condition $| \hat{O}\mathbf{R} - \mathbf{R}' | = 0$  with the tolerance factor \texttt{gen\_tol} = 1e-5. Also, applying the symmetry operations may map a given $\mathbf{R}$ to itself. All these duplicate occurrences are removed when constructing the stars. In the next step the functions $W_i(\mathbf{k})$ are constructed for $i=1,2,3,4$.

The roots of $W_i(\mathbf{k})$ functions are found in the following way. First, a $\mathbf{k}$-point grid is constructed spanning the one unit cell of the reciprocal lattice. The number of $\mathbf{k}$-points is controlled by the \texttt{nkpts} input parameter (default value 2e+7). When constructing the $\mathbf{k}$-point grid the unit vectors of the reciprocal lattice $(\mathbf{b}_1,\mathbf{b}_2,\mathbf{b}_3)$ are divided into $(n_1,n_2,n_3)$ sub-divisions so that the grid spacing is uniform. The grid spacing is computed as $\texttt{step}=[\mathbf{b}_1 \cdot (\mathbf{b}_2 \times \mathbf{b}_3) / \texttt{nkpts}]^{1/3}$ and $n_i=|\mathbf{b}_i| / \texttt{step}$. Then, the values of $W_1(\mathbf{k})$ are evaluated over the entire grid and its zeros are used as starting points near which the simultaneous roots of the $W_1(\mathbf{k})$, $W_2(\mathbf{k})$ and $W_3(\mathbf{k})$ will be evaluated. The situation in which no zeros of $W_1(\mathbf{k})$ are found for any grid point is easily remedied by increasing the \texttt{nkpts}.

In the \texttt{mvp.py} code the $\mathbf{k}$-points at which $W_1(\mathbf{k}) \leq \,$1e-12 (input parameter \texttt{zero}) are extracted and the function \texttt{fsolve} from \texttt{scipy.optimize} package is used to find roots of a vector function $\mathbf{func}_1=(W_1(\mathbf{k}),W_2(\mathbf{k}),W_3(\mathbf{k}))$ near every one of those $\mathbf{k}$-points. If the roots of $\mathbf{func}_1$ exist then the roots that minimize $W_4(\mathbf{k})$ are extracted and those that are possibly outside the reciprocal unit cell are mapped back inside. The list of these $\mathbf{k}$-points then represents the list of mean-value points (symmetry equivalent). User can then extract (on their own) those that belong to the first Brillouin zone, or choose to output only the $\mathbf{k}$-point from this set with the lowest norm (input parameter \texttt{only\_lowest\_norm}).  In case $\mathbf{func}_1$ has no roots, the roots of $\mathbf{func}_2=(W_1(\mathbf{k}),W_2(\mathbf{k}))$ are then solved for and those that minimize $W_3(\mathbf{k})$ are extracted and mapped back into the reciprocal unit cell. Lastly, our tests indicate that relatively large default value of \texttt{nkpts} is unfortunately required to converge the MVP value. Using different solvers might help converge the MVP faster, but that will be explored in the future.

The \texttt{mvp.py} code is available on github \cite{mvp:code} as part of the \texttt{toolbox} package \cite{toolbox} for \texttt{pylada} \cite{pylada:code}, a collection of python tools for structure prediction (random structure generation), surface cutting, various structure manipulations, etc. All dependencies include: \texttt{python}, \texttt{numpy}, \texttt{scipy}, \texttt{spglib}, \texttt{pylada}, and \texttt{toolbox}. The \texttt{mvp.py} is constructed around the \texttt{pylada} structure object but is easily adaptable to use other structure objects from other codes (e.g ASE \cite{ase-paper}, \texttt{aflow} \cite{curtarolo:art191}, \texttt{pymatgen} \cite{ONG2013314},\dots), which will remove the \texttt{pylada} dependency. 

Finally, the choice to write the \texttt{mvp.py} in Cartesian rather then more elegant crystal coordinates is mainly driven by the differences in how the integers are treated between \texttt{python2} and \texttt{python3}. In this way the code works with both versions. The price to pay is the introduction of various tolerance factors whose default values are set to best reflect extensive testing. That said, the  convergence of the results with respect to various tolerance factors needs to be tested before use.  
%
\section{Results}
%
Table~\ref{tab:results} below lists the MVP coordinates for the 14 Bravais lattices. For clarity, the specific unit cells that are used are displayed together with the choice of unit vectors and their coordinates. The corresponding first Brillouin zones are also displayed as well as the location of the MVP points. In the first three rows the original results of Baldereschi are reproduced. The coordinates of the MVPs for the simple cubic, face centered and body centered cubic lattices correspond well to those reported in Ref.~\cite{Baldereschi_PRB:1973}. The rest are original results. 

Importantly, coordinates of the MVPs in general depend on the choice of the lattice parameters. Table~~\ref{tab:results} only lists MVP coordinates for the specific choices provided in the table. For other lattice parameters users should compute the MVPs on their own. 
%
%
\LTcapwidth=\textwidth
\begin{longtable*}{@{\extracolsep{\fill}}lccclc}
\caption{\label{tab:results} The mean-value points (MVPs) for the 14 Bravais lattices are presented. Lattice name, the corresponding unit cell with the lattice vectors denoted, the unit cell matrix in the units of lattice constant $a$, the Brillouin zone showing the location of the MVP, the cartesian and crystal coordinates of the MVP, and finally, the values of the symmetrized waves $W_i$ ($i=1,2,3,4$) from eq.~\eqref{eq:expansion_2} at the MVP are shown. Note that this table lists results for specific choices of the lattice parameters and that MVP coordinates, both Cartesian and crystal coordinates, may be different for other values of the lattice parameters.}\\
\hline\hline
                                               &                                                                                     &                        &                        &                              & \\
                                               &                                                                                                               & lattice vectors [$a$]  &                        & MVP coordinaties \\
Lattice                                    & Unit cell                                                                       & $\begin{pmatrix} a_{1x} & a_{1y} & a_{1z} \\ a_{2x} & a_{2y} & a_{2z} \\ a_{3x} & a_{3y} & a_{3z} \end{pmatrix} $                       & \parbox[c]{6em}{Brillouin zone with MVP} & \parbox[c]{9em}{$(k_x, k_y, k_z)\,$[$2\pi/a$] $(k_1, k_2, k_3)$ [crystal] } & $\begin{pmatrix} W_1 \\ W_2 \\ W_3 \\ W_4 \end{pmatrix}$ \\
                                               &                                                                                     &                        &                        &                              & \\
\hline
                                               &                                                                                     &                        &                        &                              & \\
\parbox[c]{3em}{Simple cubic}                          & \parbox[c]{5em}{\includegraphics[width=50pt]{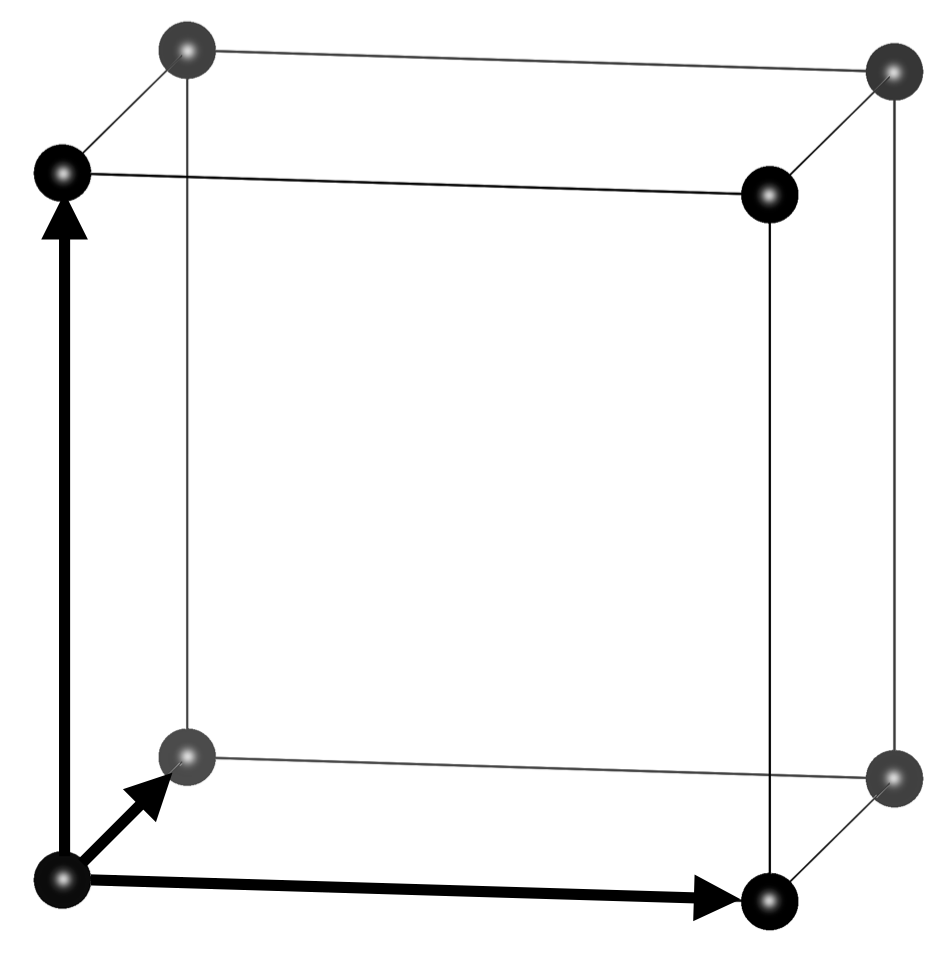}} & \parbox[c]{6em}{$\begin{pmatrix} 1. & 0. & 0. \\ 0. & 1. & 0. \\ 0. & 0. & 1. \end{pmatrix}$} & \parbox[c]{5em}{\includegraphics[width=50pt]{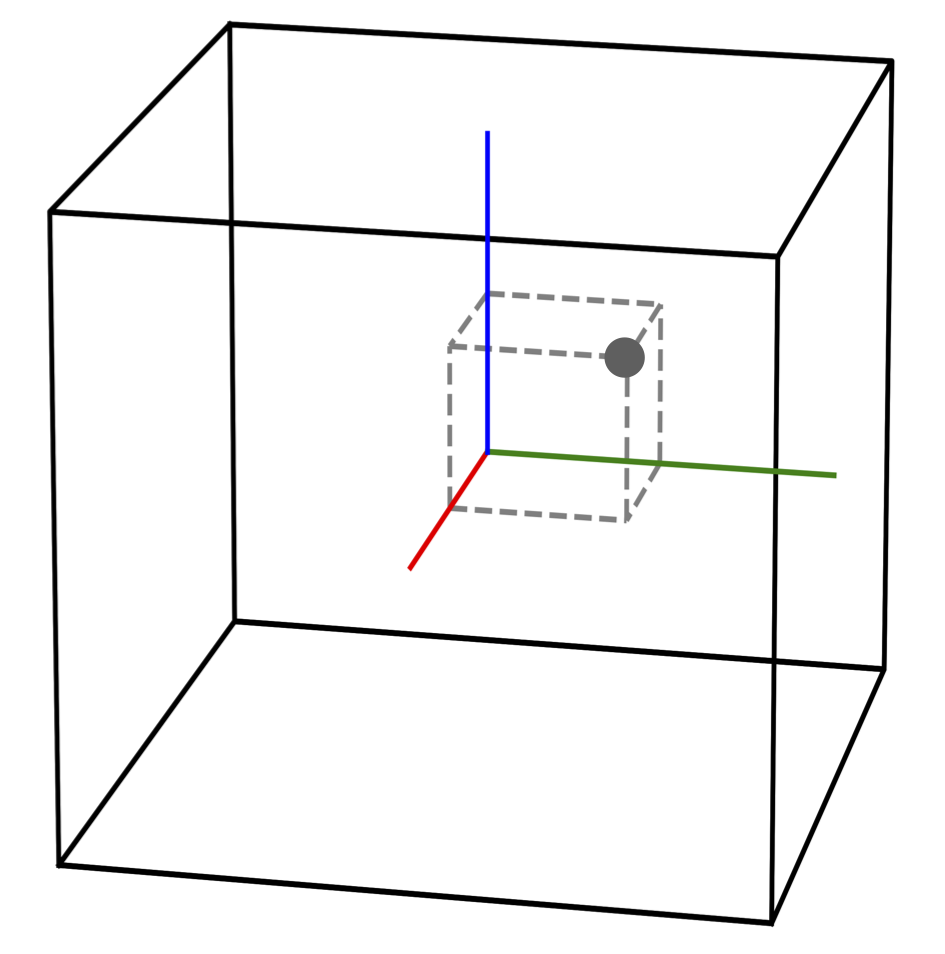}} & \parbox[c]{6em}{$(0.2500,  0.2500, 0.2500)$ $(0.2500,  0.2500, 0.2500)$}  & $\begin{pmatrix} 0.0 \\ 0.0 \\ 0.0 \\ 6.0 \end{pmatrix}$ \\
                                               &                                                                                     &                        &                        &                              & \\
\parbox[c]{4em}{Face centerred cubic}             & \parbox[c]{5em}{\includegraphics[width=50pt]{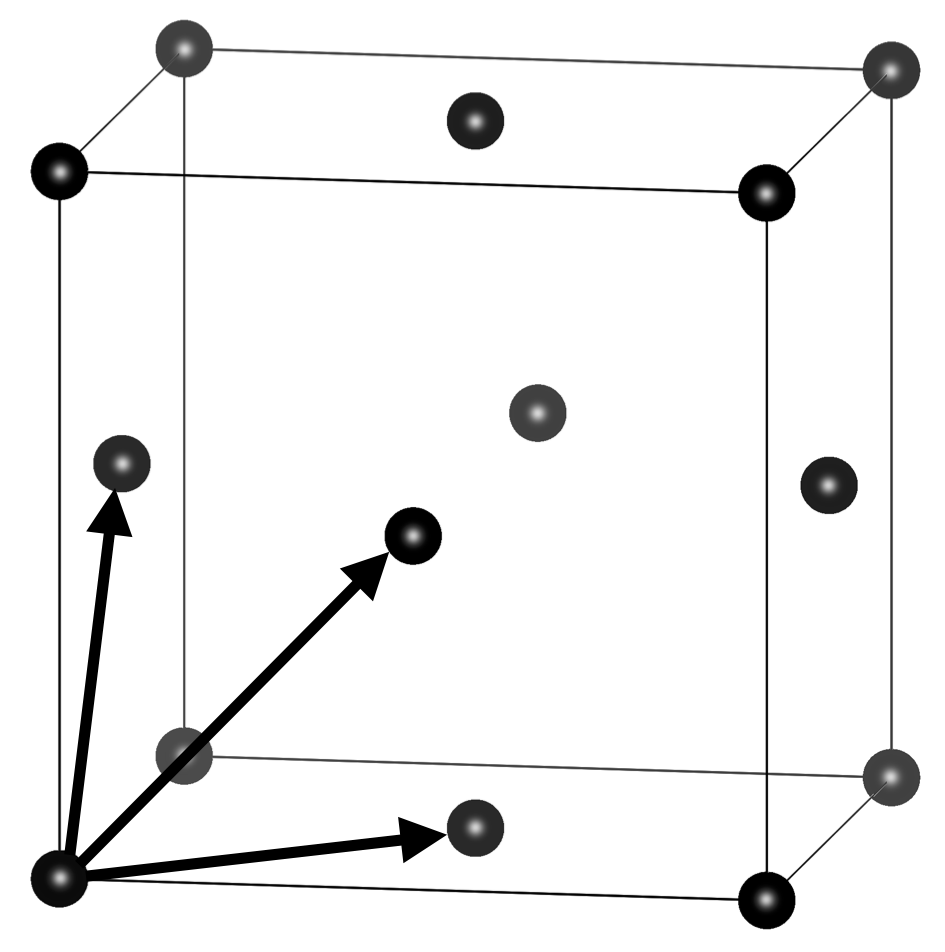}} &  \parbox[c]{7em}{$\begin{pmatrix} 0. & 0.5 & 0.5 \\ 0.5 & 0. &  0.5 \\ 0.5 & 0.5 & 0. \end{pmatrix}$} & \parbox[c]{5em}{\includegraphics[width=50pt]{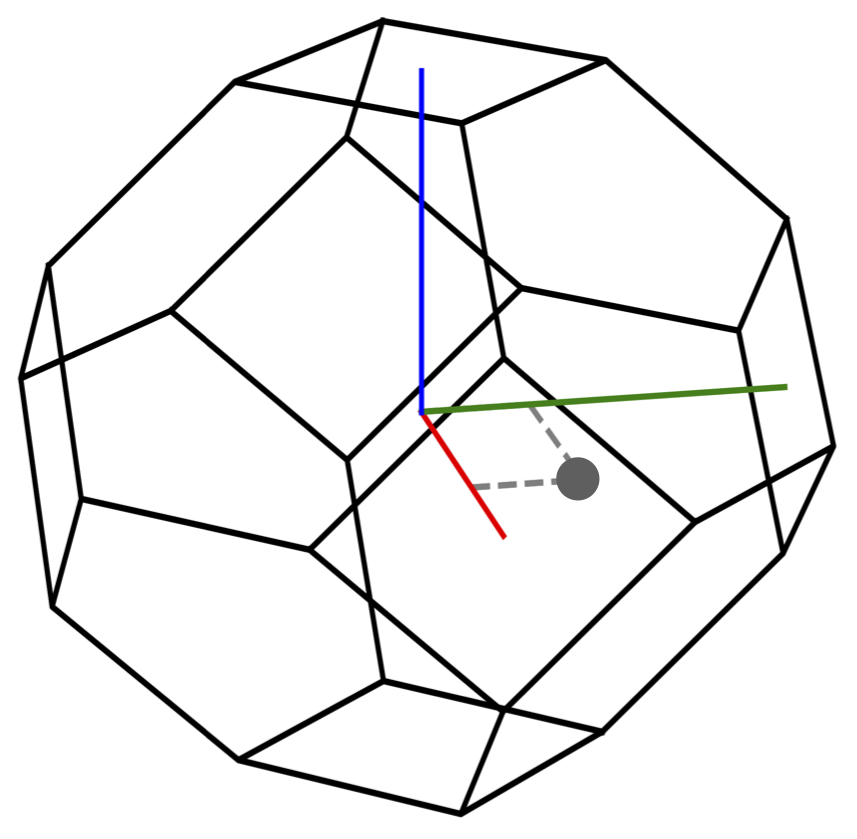}} & \parbox[c]{6em}{$(0.6223, 0.2953, 0.0000)$ $(0.1477, 0.3112 , 0.4588)$} & $\begin{pmatrix} 0.0 \\ 0.0 \\ 4.4 \\ 3.2 \end{pmatrix}$ \\
                                               &                                                                                     &                        &                        &                              & \\
\parbox[c]{4em}{Body centerred cubic}             & \parbox[c]{5em}{\includegraphics[width=50pt]{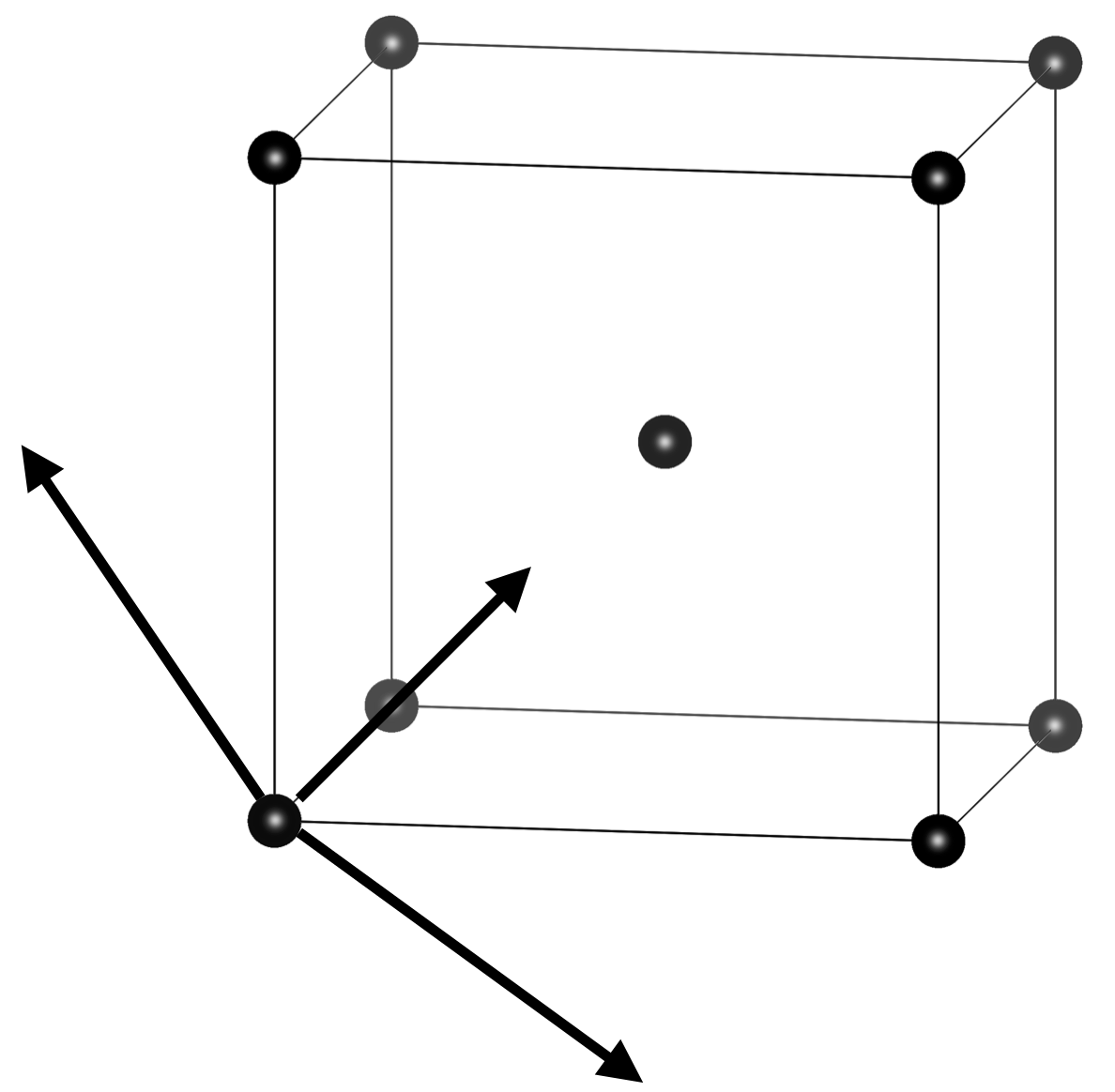}} & \parbox[c]{9em}{$\begin{pmatrix} -0.5 & 0.5 & 0.5 \\ 0.5 & -0.5 & 0.5 \\ 0.5 & 0.5 & -0.5 \end{pmatrix}$} & \parbox[c]{5em}{\includegraphics[width=50pt]{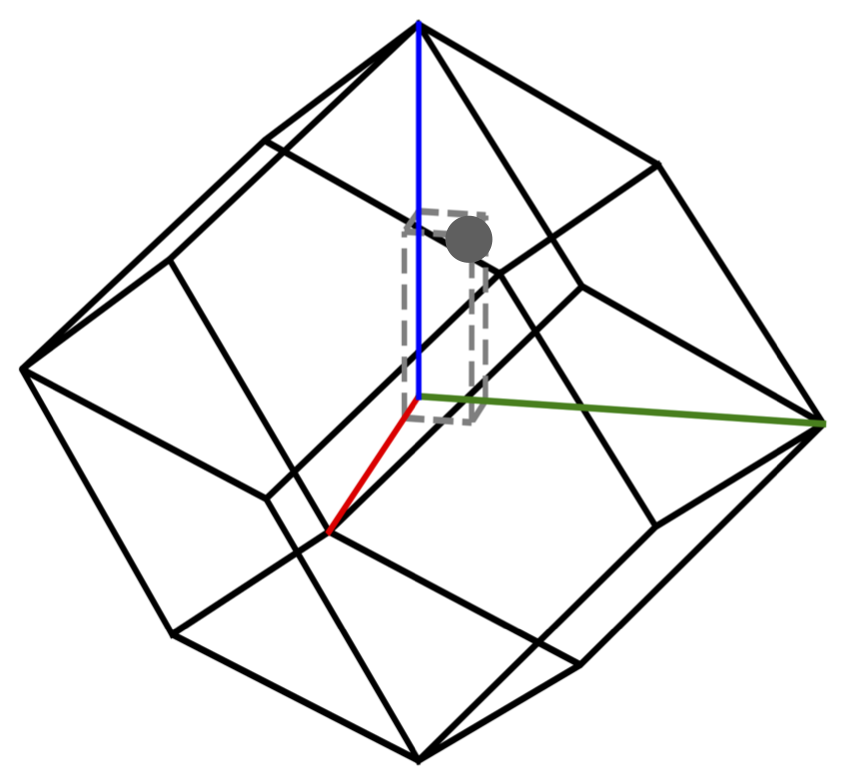}} & \parbox[c]{6em}{$(0.1667, 0.1667, 0.5000)$ $(0.2500, 0.2500, 0.9167)$} & $\begin{pmatrix} 0.0 \\ 0.0 \\ 3.0 \\ 0.0 \end{pmatrix}$ \\
                                               &                                                                                     &                        &                        &                              & \\
Hexagonal                              & \parbox[c]{5em}{\includegraphics[width=50pt]{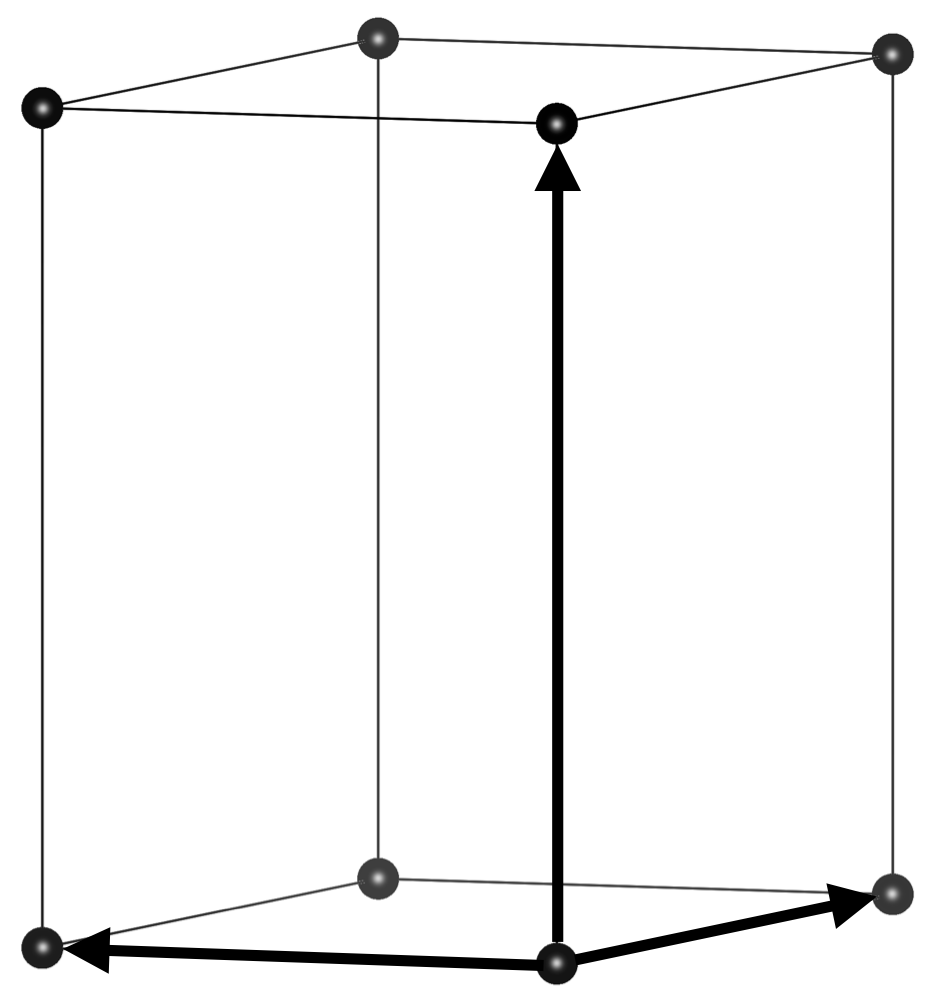}} & \parbox[c]{10em}{$\begin{pmatrix} 1. & 0. & 0. \\ -0.5 & 0.866 & 0. \\ 0. & 0. & 1.6333 \end{pmatrix}$} & \parbox[c]{5em}{\includegraphics[width=50pt]{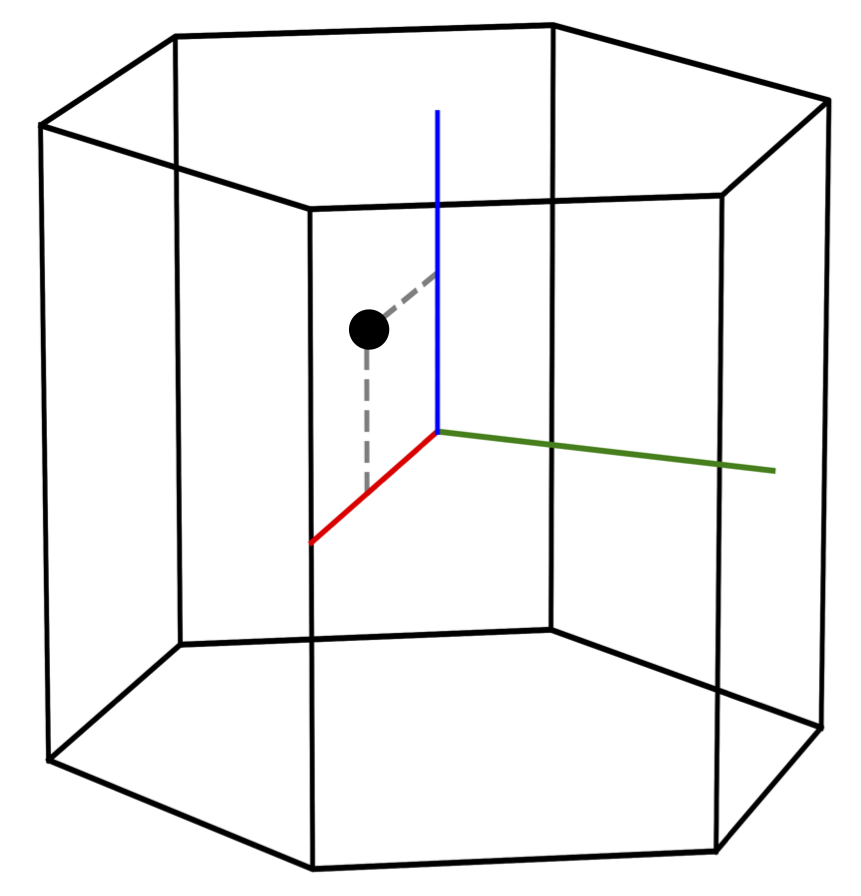}} & \parbox[c]{6em}{$(0.3807, 0.0000, 0.1531)$ $(0.3807, -0.1901, 0.2500)$} & $\begin{pmatrix} 0.0 \\ 0.0 \\ 0.0 \\ 1.6 \end{pmatrix}$ \\
                                               &                                                                                     &                        &                        &                              & \\
Rhombohedral                       & \parbox[c]{5em}{\includegraphics[width=50pt]{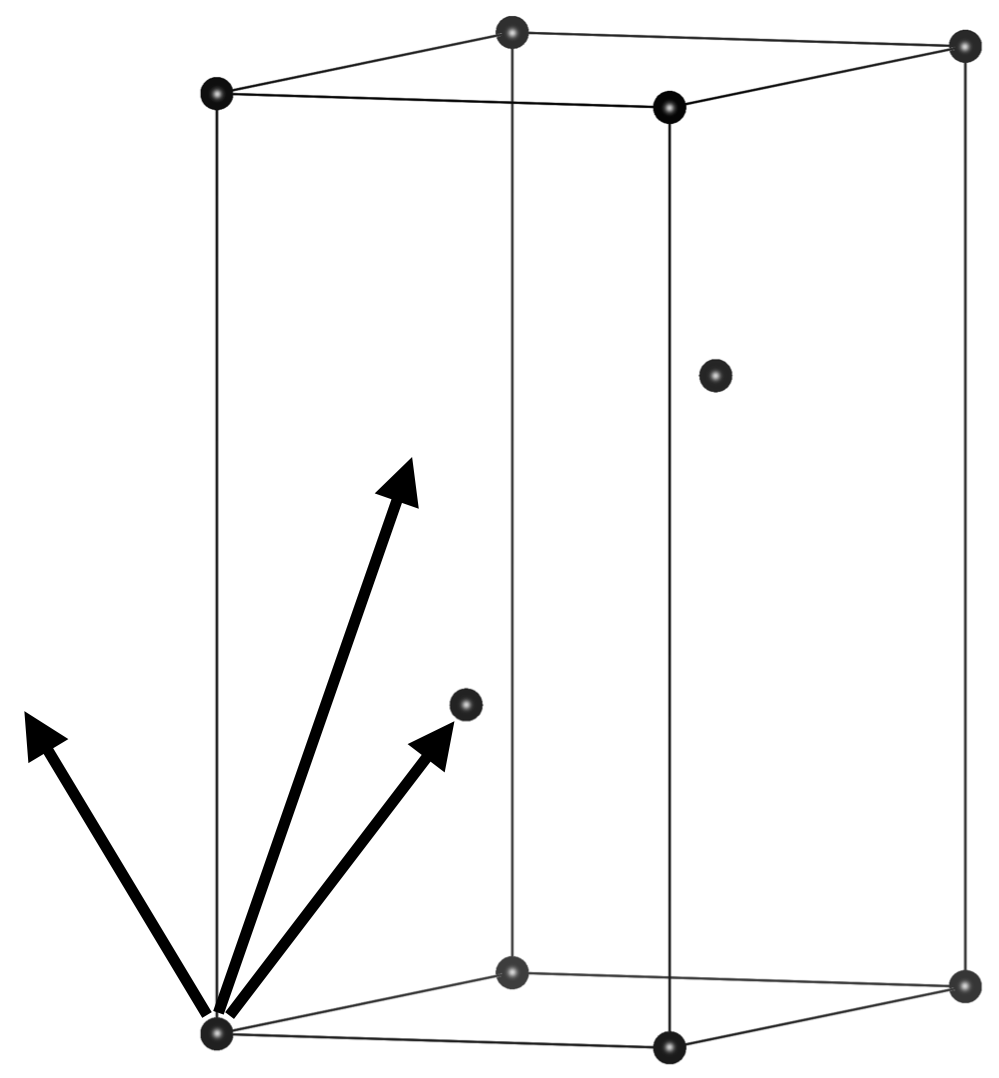}} & \parbox[c]{12em}{$\begin{pmatrix} 0.5547 & 0.3202 & 0.7679 \\ -0.5547 & 0.3202 & 0.7680 \\ 0. & -0.6405 & 0.7679 \end{pmatrix} $} & \parbox[c]{5em}{\includegraphics[width=50pt]{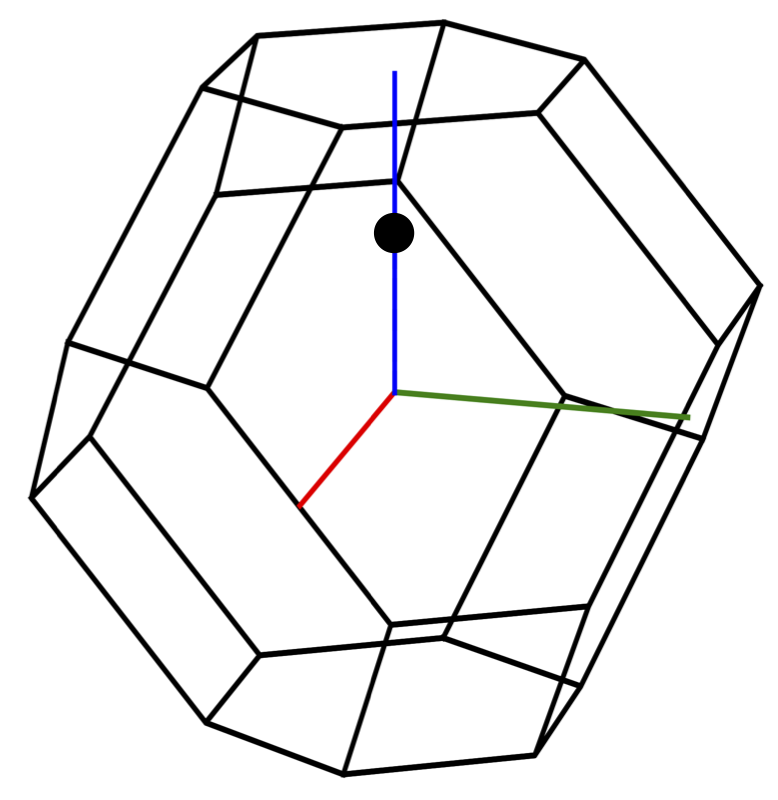}} & \parbox[c]{6em}{$(0.0000, 0.0000, 0.3255)$ $(0.2500,  0.2500,  0.2500)$} & $\begin{pmatrix} 0.0 \\ 0.0 \\ 0.0 \\ 0.0 \end{pmatrix}$ \\
                                              &                                                                                     &                        &                        &                              & \\
Tetragonal                              & \parbox[c]{5em}{\includegraphics[width=50pt]{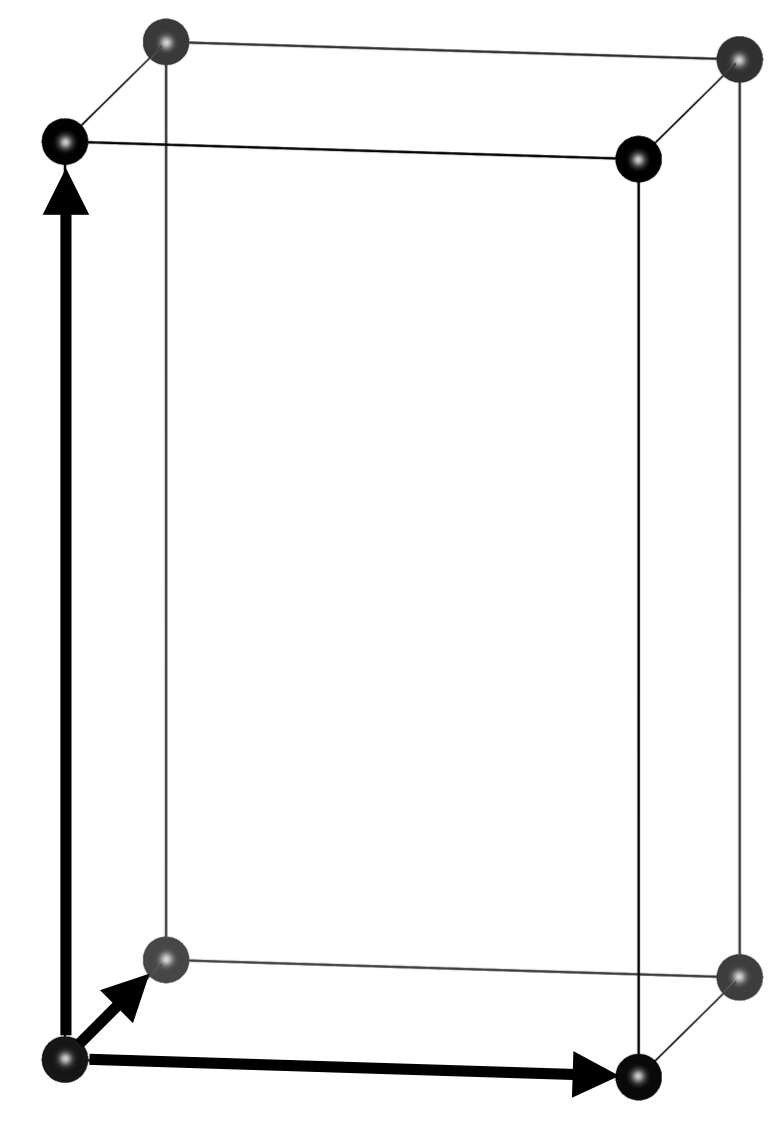}} & \parbox[c]{5em}{$\begin{pmatrix} 1. & 0. & 0. \\ 0. & 1. & 0. \\ 0. & 0 & 1.6 \end{pmatrix}$} & \parbox[c]{5em}{\includegraphics[width=50pt]{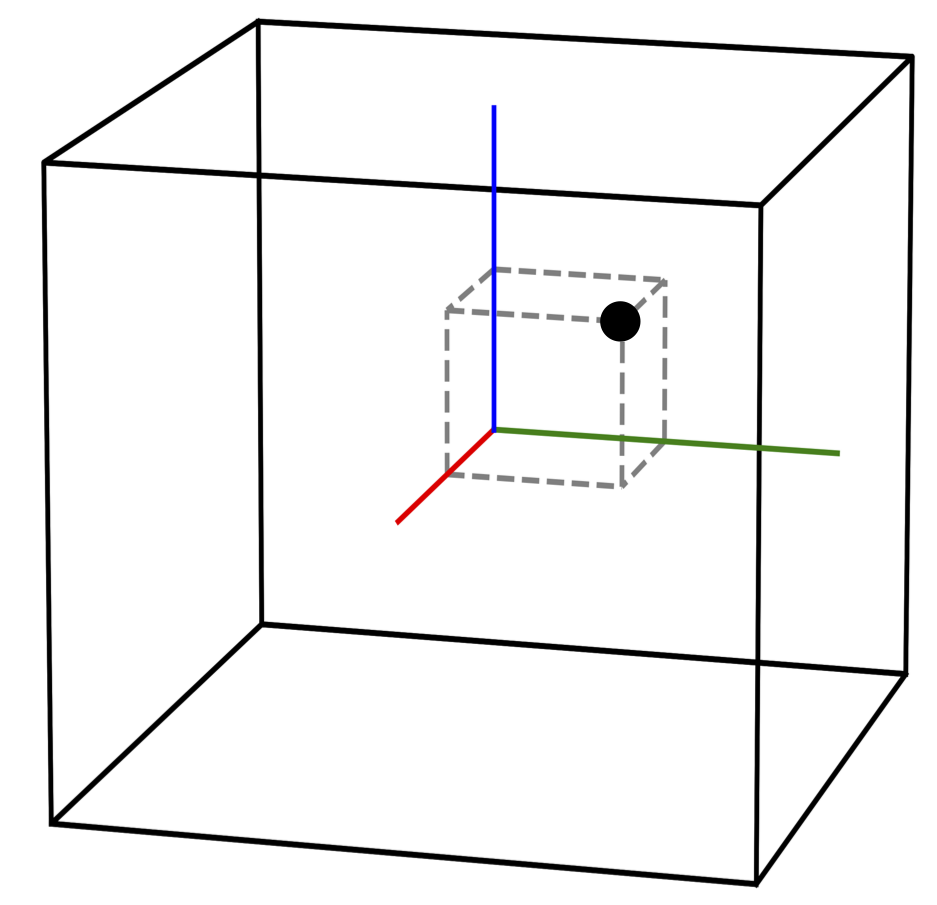}} & \parbox[c]{6em}{$(0.2500, 0.2500, 0.1562)$ $(0.2500,  0.2500,  0.2500)$} & $\begin{pmatrix} 0.0 \\ 0.0 \\ 0.0 \\ 0.0 \end{pmatrix}$ \\
                                              &                                                                                     &                        &                        &                              & \\
\parbox[c]{4em}{Tetragonal body centered}     & \parbox[c]{5em}{\includegraphics[width=50pt]{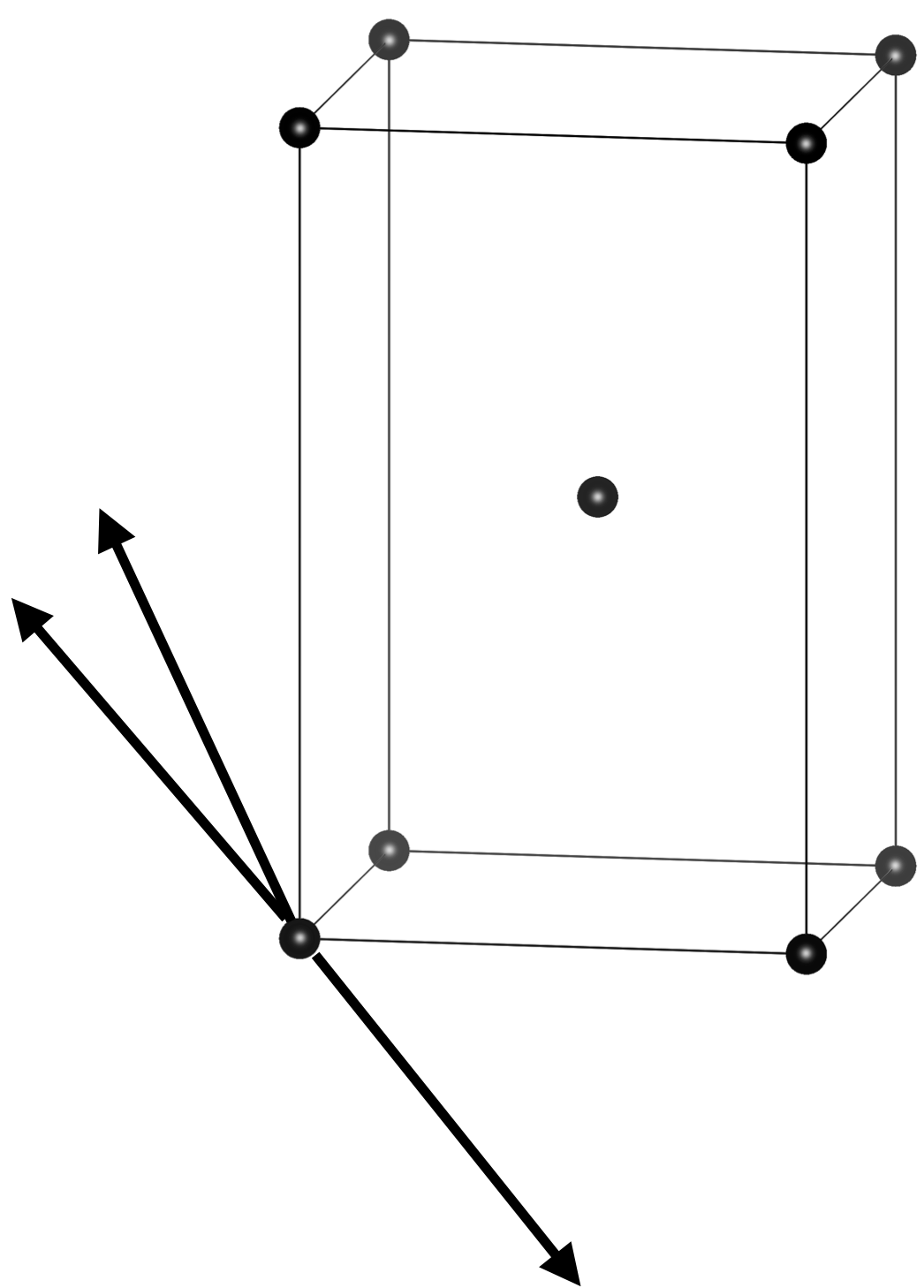}} & \parbox[c]{8em}{$\begin{pmatrix} -0.5 & 0.5 & 0.8 \\ 0.5 & -0.5 & 0.8 \\ 0.5 & 0.5 & -0.8 \end{pmatrix}$} & \parbox[c]{5em}{\includegraphics[width=50pt]{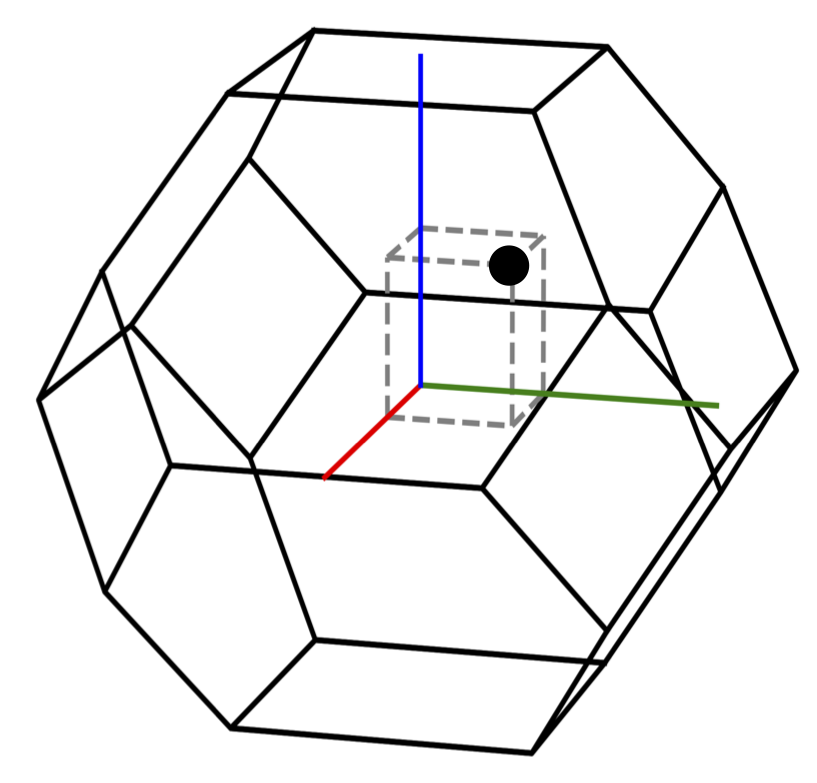}} & \parbox[c]{6em}{$(0.2500, 0.2500, 0.3125)$ $(0.2500,  0.2500,  0.0000)$} & $\begin{pmatrix} 0.0 \\ 0.0 \\ 0.0 \\ 2.0 \end{pmatrix}$ \\
                                              &                                                                                     &                        &                        &                              & \\
Orthorhombic                         & \parbox[c]{5em}{\includegraphics[width=50pt]{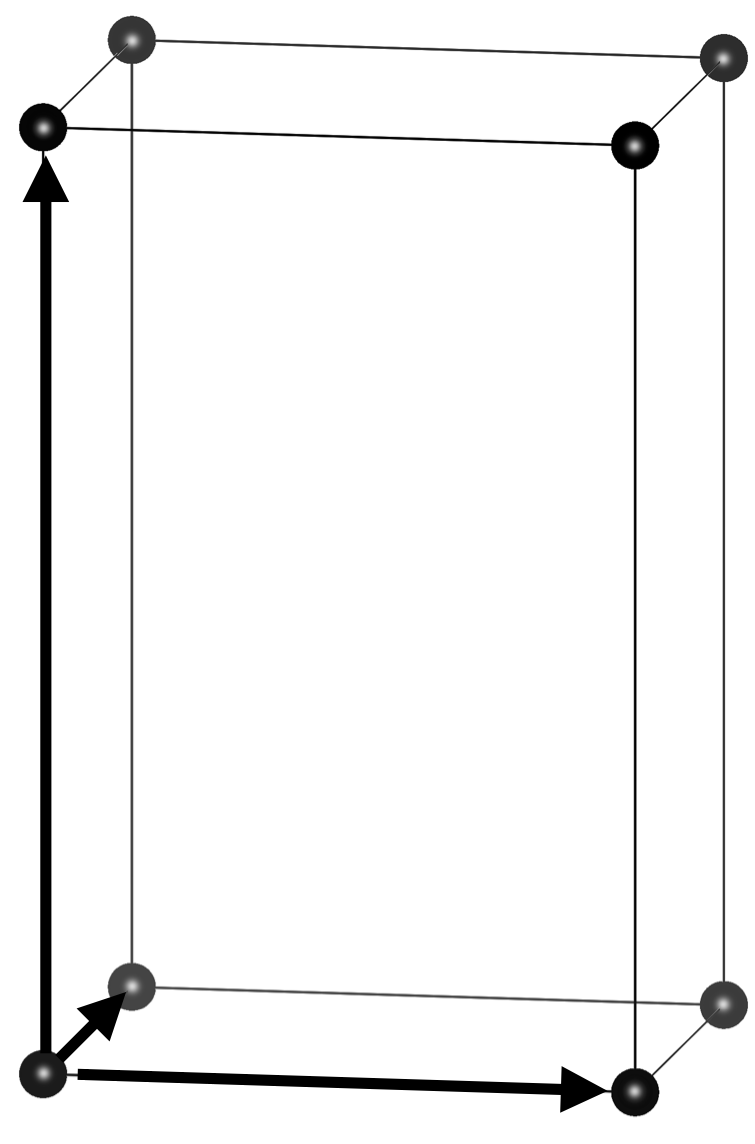}} & \parbox[c]{6em}{$\begin{pmatrix} 1. & 0. & 0. \\ 0. & 0.85 & 0. \\ 0. & 0. & 1.6 \end{pmatrix}$} & \parbox[c]{5em}{\includegraphics[width=50pt]{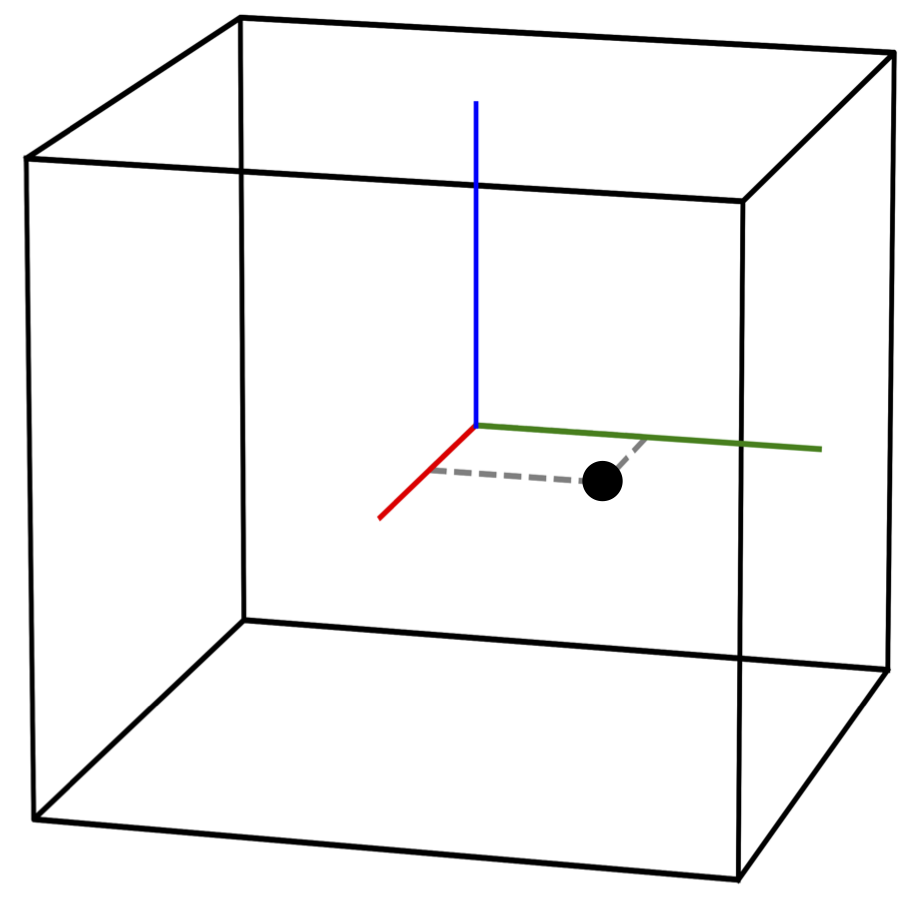}} & \parbox[c]{6em}{$(0.2500, 0.2941, 0.0000)$ $(0.2500,  0.2500,  0.0000)$} & $\begin{pmatrix} 0.0 \\ 0.0 \\ 0.0 \\ 0.0 \end{pmatrix}$ \\
                                              &                                                                                     &                        &                        &                              & \\
\parbox[c]{4em}{Orthorhombic base centered} & \parbox[c]{5em}{\includegraphics[width=50pt]{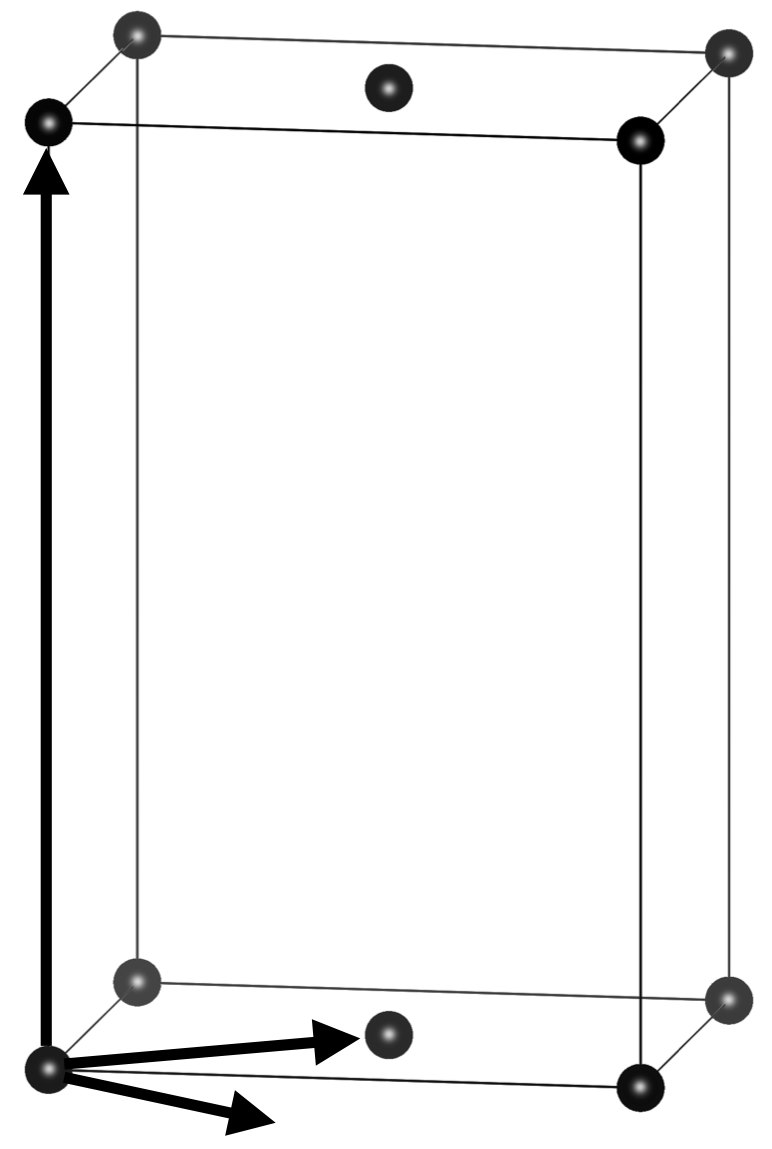}} & \parbox[c]{8em}{$\begin{pmatrix} 0.425 & -0.5  &  0. \\ 0.425 & 0.5  &  0. \\ 0.  &  0.  & 1.6  \end{pmatrix}$} & \parbox[c]{5em}{\includegraphics[width=50pt]{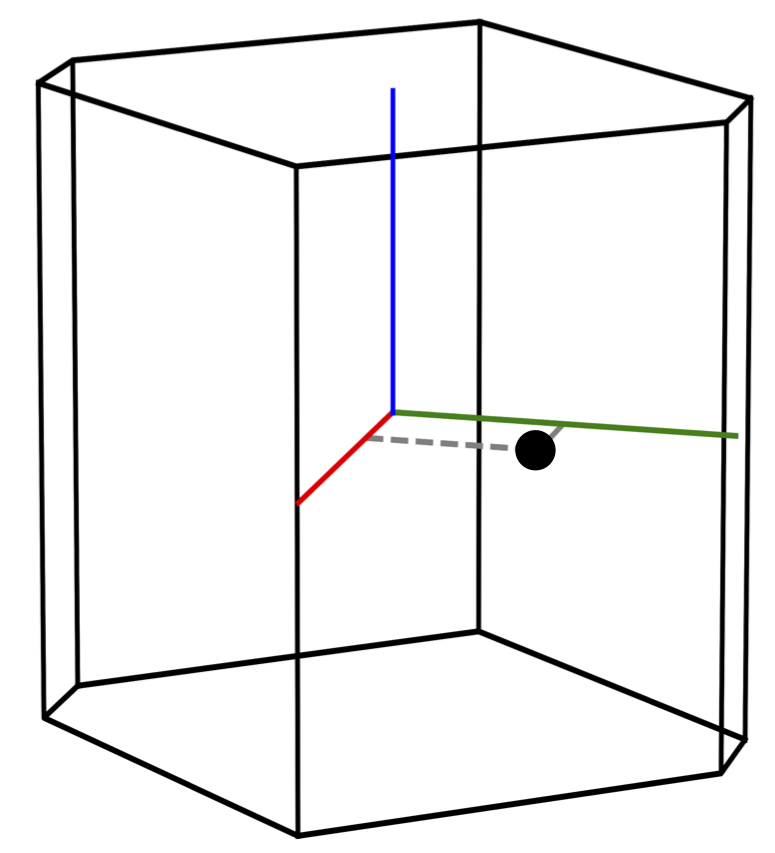}} &  \parbox[c]{6em}{$(0.2941, 0.5000, 0.0000)$ $(-0.1250,  0.3750,  0.0000)$} & $\begin{pmatrix} 0.0 \\ 0.0 \\ 2.0 \\ 0.0 \end{pmatrix}$ \\
                                              &                                                                                     &                        &                        &                              & \\
\parbox[c]{4em}{Orthorhombic body centered} & \parbox[c]{5em}{\includegraphics[width=50pt]{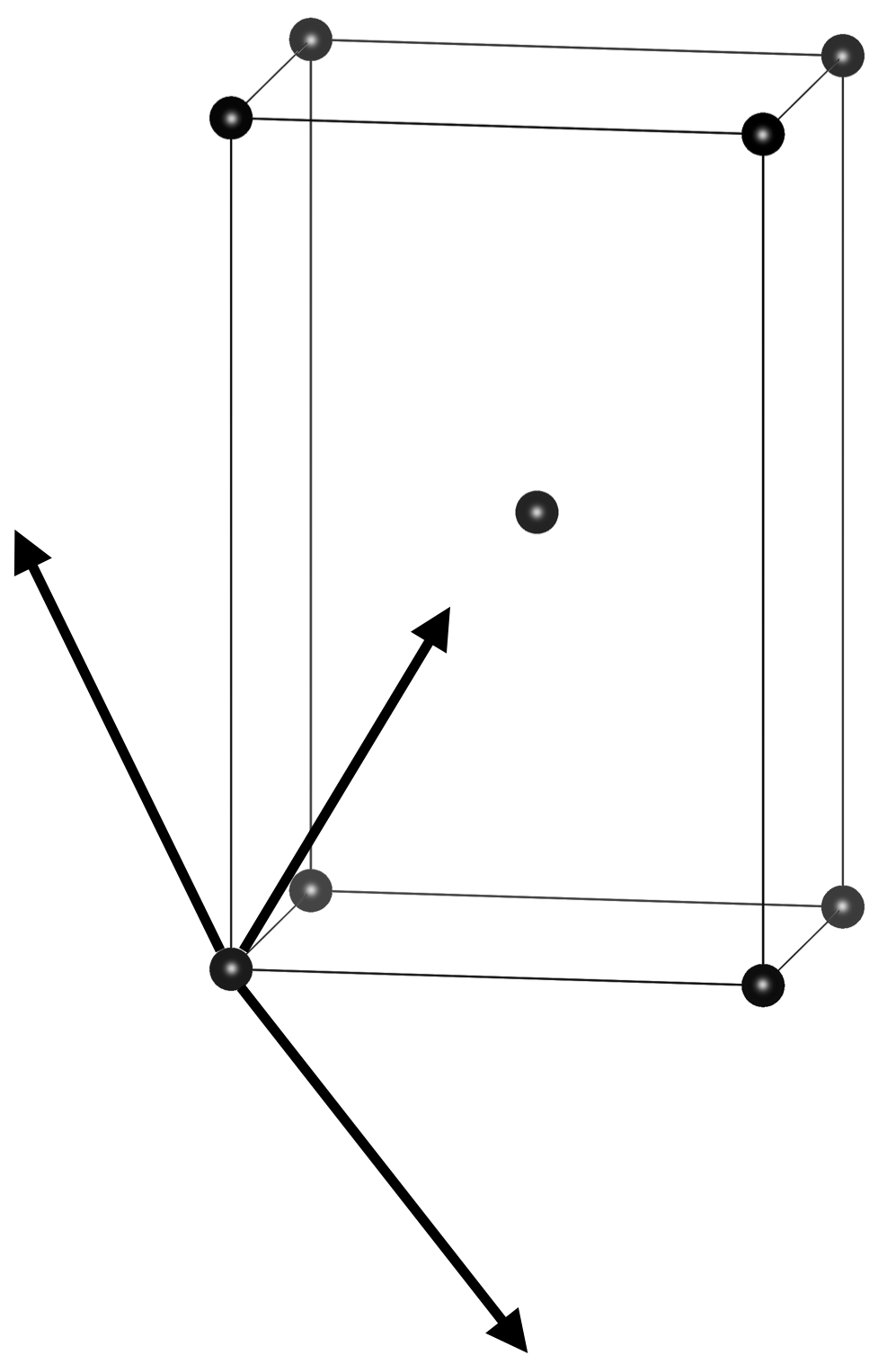}} & \parbox[c]{9.5em}{$\begin{pmatrix}  -0.425 & 0.5 & 0.8 \\ 0.425 & -0.5 & 0.8 \\ 0.425 & 0.5 & -0.8 \end{pmatrix}$} & \parbox[c]{5em}{\includegraphics[width=50pt]{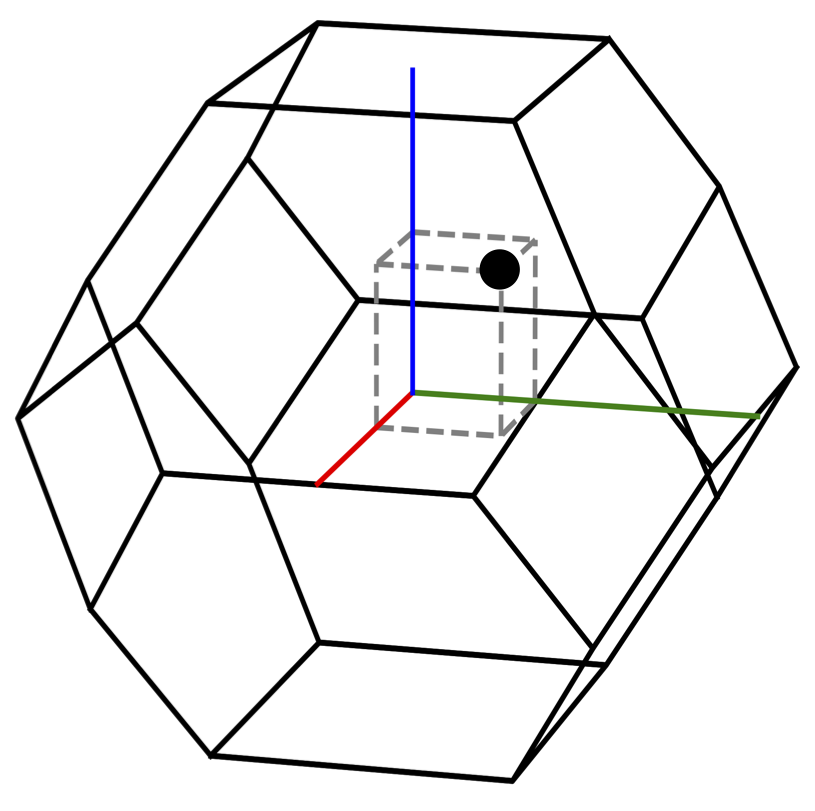}} & \parbox[c]{6em}{$(0.2941, 0.2500, 0.3125)$ $(0.2500,  0.2500,  0.0000)$} & $\begin{pmatrix} 0.0 \\ 0.0 \\ 0.0 \\ 0.0 \end{pmatrix}$ \\
                                              &                                                                                     &                        &                        &                              & \\
\parbox[c]{4em}{Orthorhombic face centered}  & \parbox[c]{5em}{\includegraphics[width=50pt]{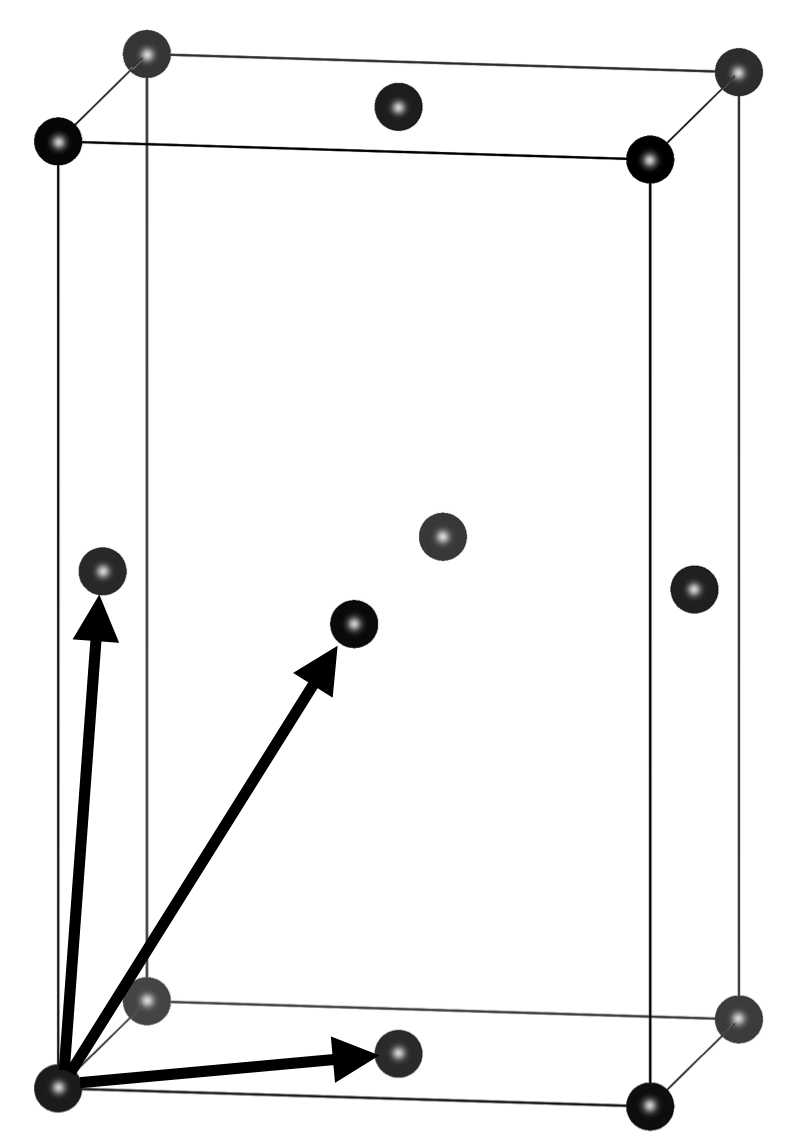}} & \parbox[c]{7em}{$\begin{pmatrix} 0. & 0.5 & 0.8 \\ 0.425 & 0.0 & 0.8 \\ 0.425 & 0.5 & 0.0  \end{pmatrix}$} & \parbox[c]{5em}{\includegraphics[width=50pt]{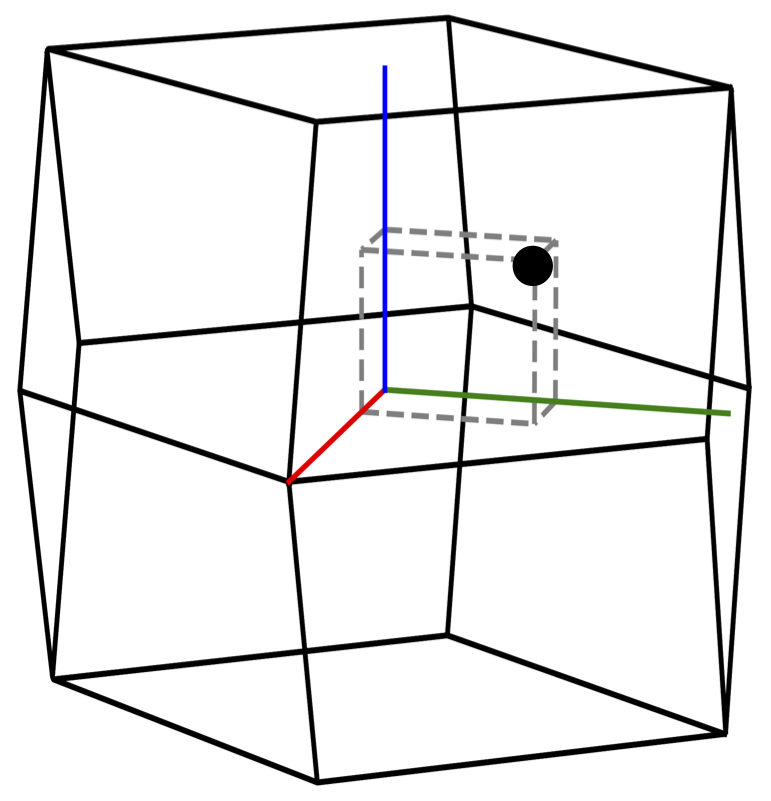}} & \parbox[c]{6em}{$(0.2941, 0.5000, 0.3125)$ $(0.5000,  0.3750,  0.3750)$} & $\begin{pmatrix} 0.0 \\ 0.0 \\ 0.0 \\ 0.0 \end{pmatrix}$ \\
                                              &                                                                                     &                        &                        &                              & \\
Monoclinic                              & \parbox[c]{5em}{\includegraphics[width=50pt]{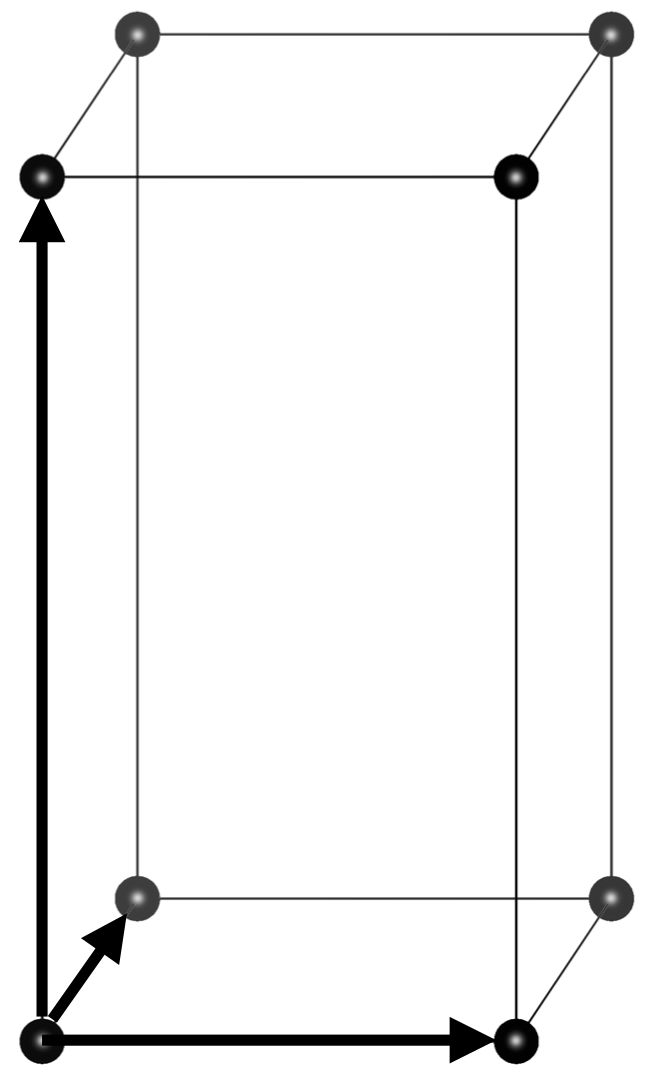}} & \parbox[c]{9em}{$\begin{pmatrix} 1.0 & 0.0 & 0.0 \\ 0.0 & 0.85 & 0.0 \\ 0.6154 & 0.0 & 1.4769 \end{pmatrix}$} & \parbox[c]{5em}{\includegraphics[width=50pt]{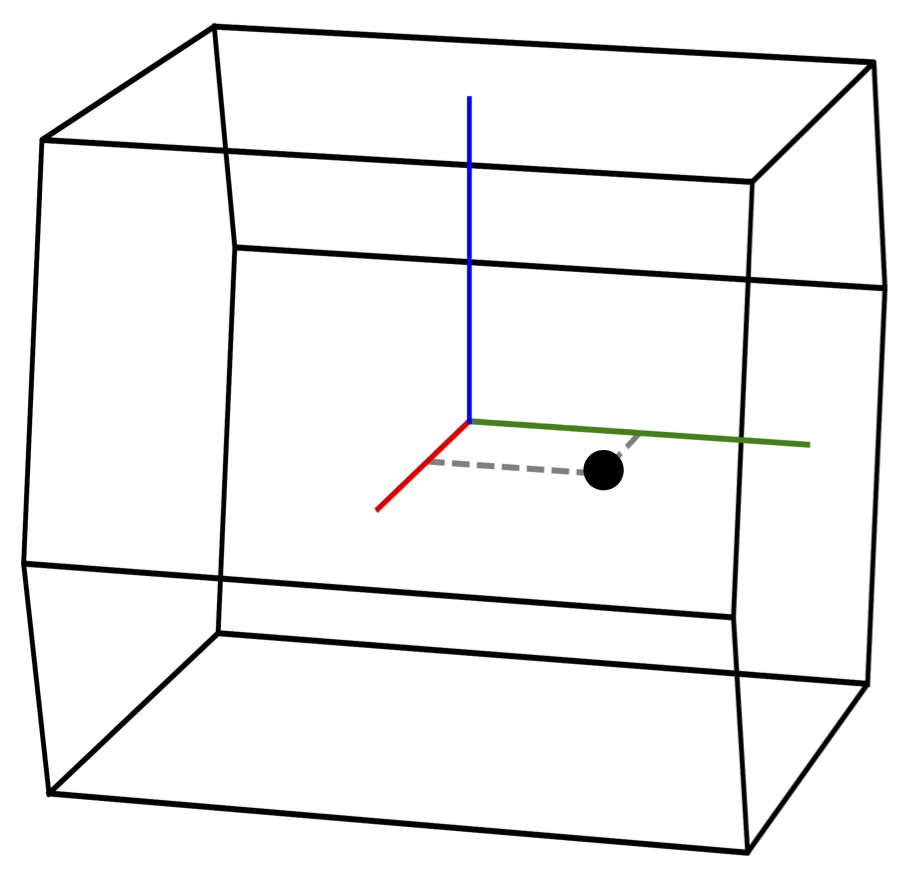}} & \parbox[c]{6em}{$(0.2500,  0.2941, 0.0000)$ $(0.2500,  0.2500,  0.1540)$} & $\begin{pmatrix} 0.0 \\ 0.0 \\ 0.0 \\ 0.0 \end{pmatrix}$ \\
                                              &                                                                                     &                        &                        &                              & \\
\parbox[c]{4em}{Monoclinic base centered}  & \parbox[c]{5em}{\includegraphics[width=50pt]{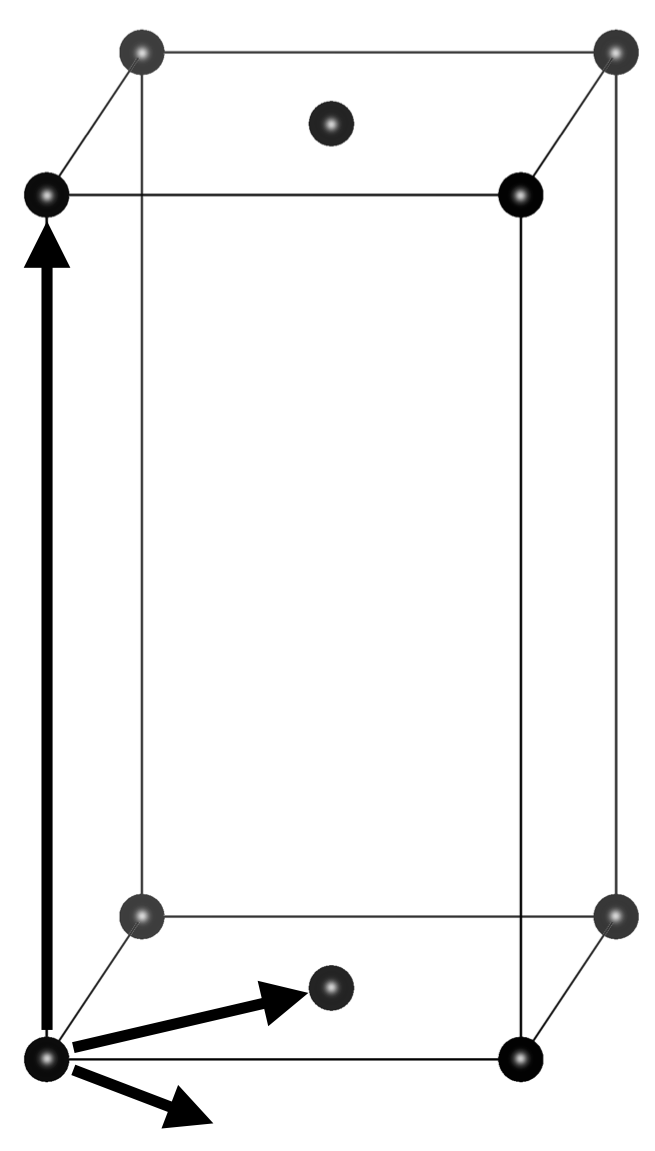}} & \parbox[c]{11em}{$\begin{pmatrix} 0.5 & -0.425 & 0.0 \\ 0.5 & 0.425 & 0.0 \\ -0.3846 & 0.0 & 1.4769 \end{pmatrix}$} & \parbox[c]{5em}{\includegraphics[width=50pt]{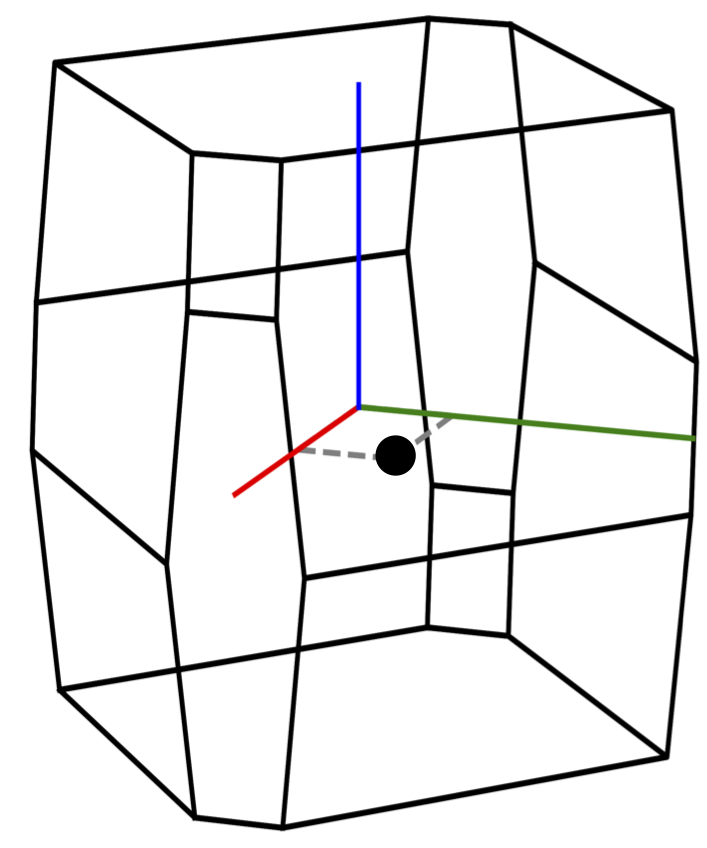}} & \parbox[c]{6em}{$(0.5000, 0.2941, 0.0000)$ $(0.1250,  0.3750, -0.1921)$} & $\begin{pmatrix} 0.0 \\ 0.0 \\ 2.0 \\ 0.0 \end{pmatrix}$ \\
                                             &                                                                                     &                        &                        &                              & \\
Triclinic                                   & \parbox[c]{5em}{\includegraphics[width=50pt]{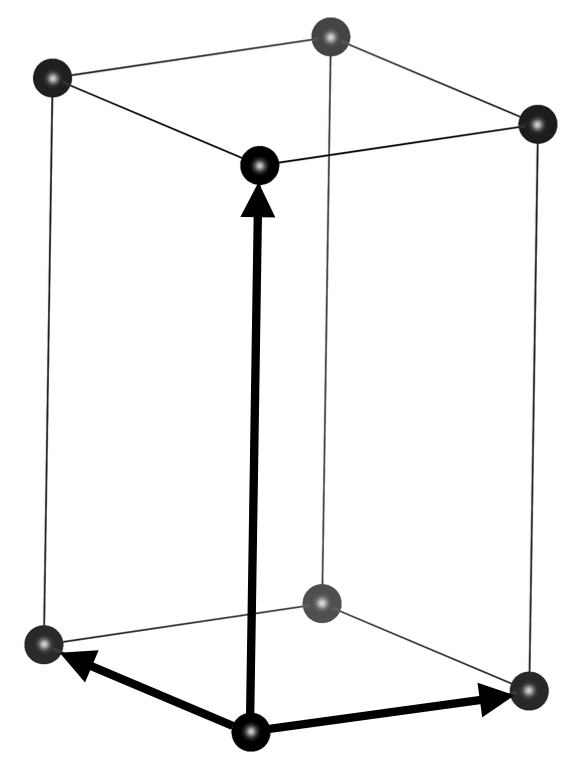}} & \parbox[c]{11em}{$\begin{pmatrix} 1. & 0.0 & 0.0 \\ 0.4682 & 0.7094 & 0.0 \\ 1.0842 & 0.0218 & 1.1764 \end{pmatrix}$} & \parbox[c]{5em}{\includegraphics[width=50pt]{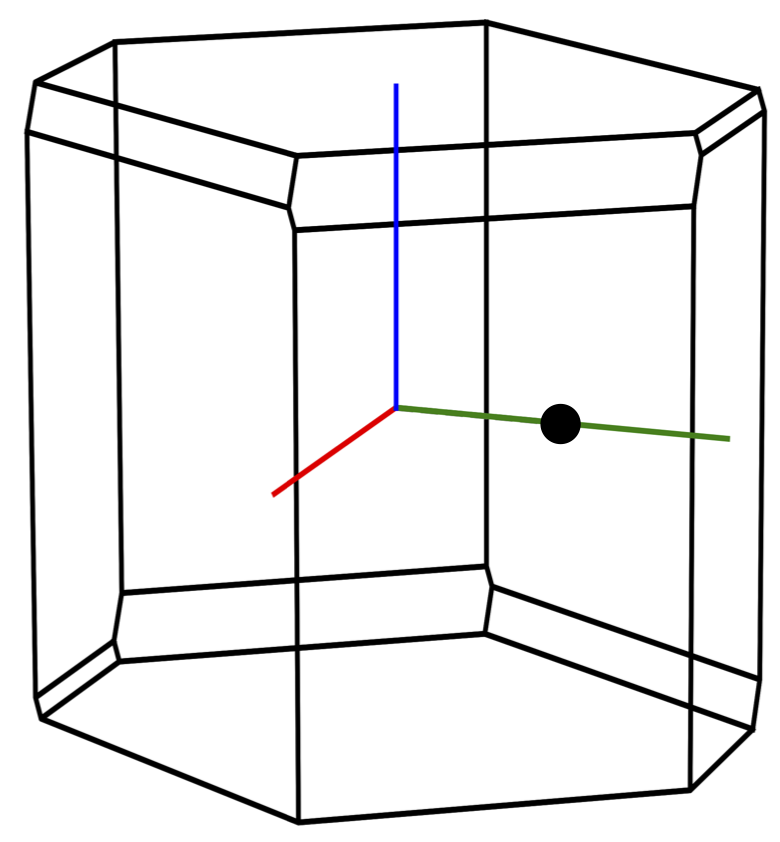}} & \parbox[c]{6em}{$(0.0000, 0.3524, 0.0000)$ $(0.0000,  0.2500,  0.0074)$} & $\begin{pmatrix} 0.0 \\ 0.0 \\ 2.0 \\ 2.0 \end{pmatrix}$ \\
                                               &                                                                                     &                        &                        &                              & \\

\hline\hline
\newpage
\end{longtable*}

%
\section{Discussion}\label{sec:discussion}
%
Here I would like to make several remarks, which may be relevant when using the code and/or understanding its structure.

It is first important to note that the systems of equations $\mathbf{func}_1=(W_1(\mathbf{k}),W_2(\mathbf{k}),W_3(\mathbf{k}))=(0,0,0)$ or $\mathbf{func}_2=(W_1(\mathbf{k}),W_2(\mathbf{k}))=(0,0)$ are non-linear in $cos(\alpha k_i)$ with $\alpha$ being an integer multiple of $2\pi$. As there is no general approach of solving such systems of equations (to my knowledge at least) the choice was to use \texttt{scipy.optimize.fsolve} function that finds roots near a certain starting point. It turns out that a relatively large number of those starting points are needed to cover most, if not all, zeros of $W_1(\mathbf{k})$, which then allows for the roots of $\mathbf{func}_1$ or $\mathbf{func}_2$ to be found. Converged values of MVPs in Table~\ref{tab:results} are obtained using \texttt{nkpts} between 1e+6 (simple cubic) and 3e+7 (hexagonal lattice). While this might look excessive, it is necessary for the  \texttt{scipy.optimize.fsolve} to find correct solutions. The typical time needed for the \texttt{mvp.py} to solve the equations (on a personal computer) is measured in minutes. 

Second, by default \texttt{mvp.py} code uses all orthogonal parts (reflections, rotations) belonging to the entire space group rather then the point group operations only. This is done so to account the fact that the entire set of orthogonal transformations may be larger than the point group alone,  and that the a group of all pure translations is invariant under the operations of all orthogonal transformations including those belongin to screw axes and glide planes. Hence, when classifying $\mathbf{R}$ vectors into stars all orthogonal transformations need to be used. While this is only relevant when dealing with real crystals (not Bravais lattices), the  \texttt{mvp.py} code is written in this way to provide access to all symmetry operations in Cartesian coordinates, which may be useful for other purposes. 

Next, the MVPs are not invariant under the supercell transformations in spite of supercells being just different representations of the same crystal/lattice. That is easy to prove just by considering the three cubic lattices. Both fcc and bcc can be represented as simple cubic using their conventional unit cells. However, the $(1/4,1/4,1/4)$ $\mathbf{k}$-point in the crystal coordinates, which is the MPV for simple cubic lattice is not an MVP for neither fcc nor bcc. This is because the first star of $\mathbf{R}$ for both fcc and bcc (set of 12 and 8 lattice vectors, respectively) is not accounted for in the simple cubic lattice and consequently $W_1 \neq 0$ for both of them at $\mathbf{k}=(1/4,1/4,1/4)$. The second star of $\mathbf{R}$ for fcc and bcc is the same as the first star for simple cubic, which then implies $W_2 = 0$ at $\mathbf{k}=(1/4,1/4,1/4)$ and so on.

Lastly, it is important to keep in mind that the MVP coordinates do in general depend on the choice of the lattice parameters. This is true for Cartesian as well as the crystal coordinates as already discussed in Ref.~\cite{Bashenov_PSSb:1977} for tetragonal and rhombohedral lattices. It is precisely this dependence that motivated writing the \texttt{mvp.py} code rather then tabulating all cases for all 14 Bravais lattices.\\
%
\section{Conclusions}
%
 In conclusion, the \texttt{mvp.py} code is presented for computing the mean-value (Baldereschi's) point in the Brillouin zone for any crystal structure. The underlying theory is also presented and discussed as are various aspects of the specific implementation in the \texttt{mvp.py} code. The code itself relies on the \texttt{pylada} python package \cite{pylada:code} a high-throughput computational physics framework, but is easily adaptable to other similar packages.
 
 \begin{acknowledgments} 
I would like to acknowledge discussions with Prof. A. Baldereschi. They were instrumental for many things I have done, this work included. The work is supported by the US National Science Foundation, Grant No. DMR-1945010.  
 \end{acknowledgments} 

%
%
%

\end{document}